\documentstyle[aas2pp4]{article}

\newcommand{\pref}{\protect\ref}

\begin{document}

\title{\large SUMER Observations Confirm the Dynamic Nature of the
Quiet Solar Outer
Atmosphere: The Inter-network Chromosphere}

\author{Mats Carlsson}
\affil{Institute of Theoretical Astrophysics, University of Oslo, 
P.O. Box 1029, Blindern, 0315 Oslo, Norway}

\author{P. G. Judge}
\affil{High Altitude Observatory,
National Center for Atmospheric Research,\altaffilmark{1}
P.O. Box 3000, Boulder CO~80307-3000, USA}

\and

\author{K. Wilhelm}
\affil{Max-Planck-Insitut f\"ur Aeronomie,
D-37189 Katlenburg-Lindau, Germany}

\altaffiltext{1}{The National Center for Atmospheric Research is
sponsored by the National Science Foundation}

\begin{abstract}

On 12 March 1996, during the commissioning phase of the SOHO mission,
we obtained observations of the quiet Sun with the SUMER instrument.
The observations were sequences of 15-20 second exposures of ultraviolet
emission line profiles and of the neighboring continua. 
These data contain
signatures of the dynamics of the solar chromosphere that are uniquely
useful because of wavelength coverage, moderate signal-to-noise ratios,
and image stability.

We focus on data for the inter-network chromosphere.
The dominant observed phenomenon is an oscillatory behavior that is
analogous to the 3 minute oscillations seen in
\ion{Ca}{2} lines. 
The oscillations appear to be coherent over 3-8 arcsecond diameter
areas.
At any time they occur over $\sim$50\% of the area studied, and they
appear as large perturbations in the intensities of lines and
continua.
The oscillations are most clearly seen in intensity variations in the
UV ($\lambda > 912$\AA) continua, and they are also seen in the
intensities and velocities of chromospheric lines of \ion{C}{1},
\ion{N}{1} and \ion{O}{1}. 
Intensity brightenings are accompanied by blueshifts of typically
5 km s$^{-1}$. Phase differences between continuum and line intensities
also indicate the presence of upward propagating waves.
The detailed behavior is different between 
different lines, sometimes showing phase lags.
Three minute intensity oscillations are occasionally seen in second
spectra (\ion{C}{2} $\lambda 1335$), but never in third spectra
(\ion{C}{3} and \ion{Si}{3}). Third spectra and \ion{He}{1} $\lambda
584$ show oscillations in velocity that are not simply related to the
3 minute oscillations.
The continuum intensity variations are consistent with recent
simulations of chromospheric dynamics (Carlsson \& Stein 1994) while
the line observations indicate that important ingredients are missing
at higher layers in the simulations.

The data show that time variations are crucial for our understanding
of the chromosphere itself and for the spectral features formed there --
the quiet Sun's chromosphere is very dynamic and not ``quiet''. The
implications of these data should be considered when planning
chromospheric work with instruments such as those on SOHO.

\end{abstract}

\keywords{Sun: chromosphere}

\section{Introduction} \label{sec:introduction}

Recent years have seen a fundamental change in our understanding of
the nature of the solar chromosphere. This has resulted from a
synthesis between observations, some aspects of which were already
\nocite{Hale+Ellerman1904}
known by Hale \& Ellerman (1904), and a specific class of
theoretical models, developed only recently to the point where
sensible comparisons with observations could be made
(Carlsson \& Stein 1994, 1995, 1997).
\nocite{Carlsson+Stein1994}
\nocite{Carlsson+Stein1995}
\nocite{Carlsson+Stein1997}
In essence, the change implies that for
regions of the chromosphere whose structure is not obviously
controlled by magnetic fields, a {\em dynamic} and not static picture
is needed to describe the structure and the emitted spectrum (see the
\nocite{Rutten1994}
review by Rutten 1994). The evidence for this is the
remarkable qualitative agreement between observed profiles of the
\ion{Ca}{2} H line and profiles computed from dynamical models in
which the equations of radiation hydrodynamics are solved for a
stratified atmosphere driven by vertical acoustic velocity
perturbations. The driving piston's behavior was set through
simultaneous observations of a photospheric absorption line. The {\em
ab-initio} model profile calculations show features that evolve with
time in a remarkably similar manner to those observed. While there
remain some discrepancies and questions, we believe that the
information content of the time-dependent \ion{Ca}{2} line profile
data, formed at heights less than 1 Mm above the photosphere, far exceeds the
information content of other data used, for example, to construct more
\nocite{Vernazza+Avrett+Loeser1981}
traditional models (e.g., Vernazza et~al. 1981).
It is clear that a static picture is very misleading, at least for the
``unmagnetized'' solar chromosphere.

The implications of this new development, if it survives further
critical analysis, are far reaching. For instance, the theoretical
models show that there is no evidence in existing observations for a
quasi-static chromosphere (Carlsson \& Stein 1994, 1995),
\nocite{Carlsson+Stein1994}
\nocite{Carlsson+Stein1995}
in contradiction with 
earlier work. Chromospheric line emission (e.g., in \ion{Ca}{2} 
lines), which in static models requires an outwardly increasing
temperature structure, is instead produced by wave motion with no
increase in the mean gas temperature. Chromospheric line absorption
in CO molecules, which in static models is inconsistent with a
1-dimensional hot chromosphere, could in principle also be produced in
the dynamical calculations. These examples show that, when dynamic
evolution of the plasma is important, the whole
foundation for using spectral features in static models to infer
physical properties of the chromosphere (in this case \ion{Ca}{2}
emission and CO absorption) must be called into question.

An important next step is to confront dynamical models with more
detailed observations. There are several pressing questions that can
be addressed with data from a space-based UV instrument like 
\nocite{Wilhelm+others1995b}
SUMER (Wilhelm et~al. 1995) on SOHO.
Can spectral features
that have varying sensitivities 
to the gas dynamics be accurately measured and 
used to get 
more information on the dynamics of the solar
chromosphere? What is the
cause of the different behavior exhibited by the \ion{Ca}{2} H line in
\nocite{Lites+Rutten+Kalkofen1993a}
the chromospheric network (Lites et~al. 1993)? 
Does the time averaged gas temperature continue to
decrease with height, as is the case in an {\em ab-initio}
one-component non-magnetic model, or increase as suggested by recently
published
\nocite{Staath+Lemaire1995}
\ion{Mg}{2} observations (Staath \& Lemaire 1995)? Is
there a connection between the dynamics in network regions deep
down and in inter-network regions higher up? What can we learn from
studying the time dependence of other emission lines whose formation
is poorly understood (e.g., lines of \ion{He}{1}). 
Are the influences of seeing fully understood in ground-based
data?

To begin to answer these questions we have initiated observing
sequences with the primary emphasis of obtaining time-series of
selected line profiles and continua with SUMER. In this paper we
discuss observations of chromospheric spectral features
obtained with SUMER during the commissioning phase of SOHO. These
cover some network regions as well as regions that appear to share
profile variations qualititively similar to the \ion{Ca}{2} data
seen in the non-magnetic chromosphere. While full analysis of these
data must await the results of radiation hydrodynamical simulations
(under way), there are some important results that do not depend on
such work, and these are reported here. An accompanying paper
discusses the network behavior, the \ion{He}{1} 584\AA\ resonance
line, and lines formed in the solar transition region.

\section{Observations and Data Reduction}

SUMER is a normal incidence spectrograph that operates between
wavelengths of 660 and 1600\AA\ (first order, and half this, second
order). Areas of the Sun are imaged by the primary mirror 
onto slits of various sizes
before the solar light is passed to the spectrograph. A portion of
the diffracted light is then imaged onto a crossed delay line
microchannel plate detector of size $360\times 1024$ pixels.
Roughly 40\AA\ (in first order) of the solar spectrum can be placed on
the detector at any given time. The spatial resolution is $\sim
1\times 2$ arcseconds (1 arcsecond slit width, 1 arcsecond sampling along
the slit). A spectral resolution element (pixel size) is $\sim
40$ m\AA\ (first order), $\sim 20$ m\AA\ (second order).

The sequence of SUMER observations discussed here was obtained on 12
March 1996 (see table~\ref{tab:observations}). The $1\times120$
arcsecond slit was illuminated by a region of the quiet Sun close to
disk center. Immediately prior to the observing run a
spectroheliogram was obtained in the \ion{O}{4} 
resonance lines at 790\AA\ 
as part of the commissioning activity. The initial slit position cuts
through several network arches but no plage. The most likely
``typical'' inter-network region, i.e. free from obvious strong
transition region emission, lies between slit positions 90 and 105.

The total time taken for the time-series observations was about 4
hours with one hour for each of the four wavelength regions. For
these particular observations no supporting observations were made
from the ground or other instruments on SOHO. Unfortunately solar
tracking, i.e. compensation for solar rotation, was active only
between the four one hour time sequences, not during the time
sequences. Thus, over each 1 hour time-series the slit mapped out the
same $10\times120$ arcsecond$^2$ area. This must be kept in mind
since without compensation, a completely new area of the Sun at disk
center is rotated onto the 1 arcsecond wide spectrograph slit every
\nocite{Allen1973}
$t_{new}=383$s (using mean solar data from Allen 1973).
Thus, we can only examine variations for a given area of the Sun on
timescales substantially less than this, and we can examine
differences between areas of different longitudes on timescales longer
than this.

Grating positions were chosen to obtain profiles of lines at
wavelengths listed in table \ref{tab:observations}. The exposure time
used was 15 or 20 s, chosen to fit the telemetry constraints for
the SUMER instrument (10.5 kbits/s) after which the detector (detector
A) was read and prepared for the next exposure. Owing to telemetry
constraints, a window of just $50$ (spectral) by 120 (spatial) pixels
was transmitted to the ground for each of the wavelength settings listed in
table \ref{tab:observations}.

\subsection{Data reduction}

The raw data consist of time-series of ``images'' with wavelength on
the x-axis and distance along the slit along the y-axis, one image
for each exposure. Each image consists of raw counts per ``pixel'' on
the detector. The sequence of 240 or 180 such images forms the ``data
cube''. 

In the present paper we wish to study line profiles
and continuum intensities as functions of time. Our data reduction
requirements are thus to obtain line profiles with high relative
photometric precision. Thus, we must reliably perform the following
tasks: flat field corrections to account for pixel-to-pixel
sensitivity variations; geometric corrections to account for
distortion of the image of the slit projected onto the Sun in both the
spectral and spatial directions; wavelength calibrations.
To relate observed intensities to simulations we also need a
radiometric calibration.

High signal to noise flat-field exposures contribute significantly to
the degradation of the detector and are therefore not taken more often
than about once per month. The flat-field exposure closest to our
observing run of 12 March was taken 28 February. Unfortunately it was
found that the flat-field pattern of sensitivity variations
($\pm50$\%, peak-to-peak) was slightly displaced in the observations compared with
the flat-field exposure. The following procedure was employed to
obtain an optimum removal of pixel-to-pixel sensitivity variations:
The mean images over the one hour observing sequence in each line was
used as a starting point to fix co-alignment of the flat to the
solar data. Each column was divided by the column mean (removing the
spectral lines) and each row was thereafter divided by the row mean
(removing intensity variations along the slit). The remaining image
contains the small scale sensitivity variation pattern. This image was
cross-correlated with the flat-field image to determine an optimum
shift of the flat-field. Such a shift was determined individually for
each line and the shifted flat-field was used subsequently. The
maximum shift determined was 0.6 pixels in the wavelength direction
and 0.9 pixels in the slit direction.
In addition to the pixel-to-pixel variation in the flatfield there is
an artifact from the AD-converter causing every second row to get 
higher counts. This effect does not shift in position with time and
was therefore taken out before the shift of the flat-field and 
multiplied in again after the shift.
The flat-fielded images were corrected for geometrical distortion
using data from T. Moran (private communication).
The wavelength calibration was taken from the pre-flight calibration
and is very inaccurate. Therefore, Doppler shifts discussed here
are relative to the mean observed wavelength of the line.
\nocite{radiometry1997}
The radiometric calibration was taken from Wilhelm et~al. (1997).

\section{Results}

We have examined the SUMER data using different methods, including
Fourier spectra and various slices through and moments of the data
cube, such as mean intensities, velocities, linewidths. The simplest
and most illuminating way to see the gross properties of the data is
to look at moments. More detailed analyses using power spectra or
wavelets will be discussed in future papers.

First consider Fig.~\ref{fig:fig_qs} which shows a slice through the
data cube of one time-series at a single position along the slit. It
shows the intensity as a function of wavelength and time for the three
lines \ion{N}{1} 1319, \ion{C}{1} 1329, \ion{C}{2} 1334 (lower panel)
and \ion{O}{1} 1355, \ion{O}{1} 1358, \ion{C}{1} 1364 (upper panel),
at a typical inter-network slit position ($x=95$). Each image is 
shown on a linear scale, each individually scaled. The average peak
count rates are 20, 7 and 8 counts pixel$^{-1}$ in the 15 s
integration times for the three upper panels, and 20, 6 and 100 counts
pixel$^{-1}$ for the lower panels respectively. The scaling makes
the continuum most visible in the upper panels and lower middle panel
although the continuum signal is similar at all wavelengths (typically
1 count pixel$^{-1}$ in 15 s). Continuum brightenings are very
evident as horizontal bands. The time between two
brightenings is typically 200s. 
The brightenings are also seen
in the line emission, especially in the
\ion{N}{1} and \ion{C}{1} lines but the brightest ones also in
\ion{C}{2} (e.g., at times t=670, 2450 and 3100). The brightenings are
always of longer duration (full widths at half maximum intensity are
$\sim 100$ s) in the \ion{N}{1} line than in the two other lines
($\sim 50$ s). The maximum intensity typically occurs first in
the continuum and simultaneously in \ion{C}{1}, about 12s later
in \ion{C}{2} and 25s later in \ion{N}{1}. The time delays vary
from grain to grain but are of similar magnitudes.

Next consider wavelength\--integrated quantities: the total line
intensity $I_{tot}$ and the mean velocity shift $\overline v$
computed from:

\begin{eqnarray}
\label{eqn:itot}
I_{tot} = \int_{\Delta \lambda} \left ( I_\lambda - I_{cont} \right )
\ d \lambda \\
\label{eqn:v}
\overline v = \int_{\Delta \lambda} \left (I_\lambda - I_{cont} \right
) \ v_{\lambda}\ d \lambda / I_{tot}
\end{eqnarray}

\noindent where $I_{cont}$ is a background continuum intensity.
All intensity data are given in counts/exposure on the detector.
$\overline v$ is given in Doppler units of
km~s$^{-1}$, and $v_\lambda = c \cdot {{\lambda - \lambda_0} \over
{\lambda_0}}$ where 
$\lambda_0$ is 
the rest wavelength of line center.

Fig.~\ref{fig:fig_moments_qs} shows
$I_{cont}$, $I_{tot}$ and $\overline v$ 
 as a function of spatial
position and time for the \ion{N}{1} 1319 and \ion{C}{2} 1334 lines.
Only a subset (spatial pixels $x=$ 65 to 120) of slit positions are shown
to highlight inter-network regions. 
The intensity data ($I_{cont}$ and $I_{tot}$) reveal the spatial
extent along the slit and the omni-presence of the bright grains. In the
inter-network region ($x$= 90-105) there are intensity
brightenings with 3-8 arcsecond spatial scale (along the slit) at all
slit positions and in all the neutral lines. The
\ion{C}{2} 1334 line is qualitatively different but also shows bright
grains correlated in time with the grains in the continuum and the
neutral lines.

The continuum intensity brightens by up to a factor of seven. The
radiation temperature at 1300 {\AA} varies between 4400\,K and 5000\,K
with an rms variation of 86\,K. This is consistent with the simulations
\nocite{Carlsson+Stein1994}
by Carlsson \& Stein (1994).

On timescales longer than $t_{new}=383$s (6.4 minutes), the typical
number of repetitions of grains at each $x$ position seen in
Fig.~\ref{fig:fig_moments_qs} can in principle be used to set limits
on the spatial and/or temporal properties of the grains. Grains are typically seen in
vertical strings (i.e., the same $x$ position in plots similar to
Fig.~\ref{fig:fig_moments_qs}) for between 15 and 30 minutes, but
can also be seen just individually and up to the full observation time
of 1 hour. 
The durations in time and widths of the grains seen along the
slit are consistent with the grains having
a diameter of a few arcseconds. It is not
possible to determine the
``lifetime'' over which a region generates grains from this dataset.

The velocity data ($\overline v$) show
interesting properties. The bright grains are seen as regions of
blueshifted emission (see, e.g., $x=95$ at
time $t=670$s in Fig.~\ref{fig:fig_qs} and \ref{fig:fig_moments_qs}). The \ion{C}{2}
$\overline v$ data reveal a remarkable oscillatory behavior that
consists of 5-15 arcsecond long oscillatory striations of peak-to-peak
amplitude $\pm 2-3$ km~s$^{-1}$. These appear to be associated with
the grains, as seen in Fig.~\ref{fig:fig_moments_qs} and
\ref{fig:fig_timeser_qs}. 
Note, however, that the velocity signal
appears coherent over larger areas, especially in regions of intermediate
line intensity (see, e.g., $x=80-90$ in Fig.~\ref{fig:fig_moments_qs}).
The $\overline v$ data also show horizontal propagation (inclined
structures in Fig.~\ref{fig:fig_moments_qs}). 
Oscillatory behavior in $\overline v$ of \ion{He}{1} and \ion{Si}{3} lines
is common, but not simply correlated with the underlying
cell flashes.

Several other general properties of grains emerge when the above data
are considered with the other wavelength regions in our data set. All
chromospheric lines show emission above the continuum everywhere, all
of the time. The \ion{Ca}{2} cell flash phenomena are seen in all
lines of neutral C, N, and O and all continua. Continuum intensities
show most clearly the signature of the grains: typical behavior is
seen in Fig.~\ref{fig:fig_moments_qs}, where the grains are seen as
flat-bottomed brightenings (sudden brightenings on timescales down to
the 15s integration times over several spatial pixels) followed
by a decay in brightness that appears ``fuzzy'', on timescales of
$10^2$s or so. Different lines within the same atom or ion can show
rather different time behavior, for example \ion{N}{1} 1319\AA\ shows
qualitatively a very different behavior from 
\ion{N}{1} 1199\AA. Although obtained at different times 
(table~\pref{tab:observations}), this behavior emphasizes the need
for
radiation (magneto-)
hydrodynamic modeling.

Grains can be seen in second spectra (\ion{C}{2}), but not in third
spectra (\ion{Si}{3} or \ion{C}{3}). 
Measured line shifts
(bulk fluid velocities determined from the first and zeroth wavelength
moments of the intensities) typically yield a $5$ km~s$^{-1}$
blueshift during the bright phase of the grains.

\subsection{Conclusions}

Our main emphasis has been to present the qualitative behavior of UV
lines and continua in the inter-network chromosphere as observed with
the SUMER instrument, point out the salient features, and draw some
preliminary conclusions. While radiation hydrodynamic calculations
are needed for detailed interpretation of these data, there are
some conclusions that can be drawn and we can speculate on others.

We can conclude that the grains appear to be 3-8 arcsecond diameter
blobs. Thus, the photospheric p-modes, with whatever controls their
upward propagation into the chromosphere, apparently provide a
coherent driver over this area to produce observable grains of this
size (this will be contrasted with the case of network time-series data
in the following paper).
The grains are extremely common, covering typically 50\% of the
observed area at a given time.

The non-detection of any grain oscillations in third spectra (\ion{C}{3} and
\ion{Si}{3}, traditionally classified as ``transition region lines''),
indicates that the upward propagating shocks that are assumed to be
responsible for the oscillations seen in the other lines and continua
are not responsible for the heating of the lower transition region.
This is discussed in the following paper.

The {\em continuum} intensity variations are consistent with the simulations
by Carlsson \& Stein (1994, 1995, 1997). However, the simulations
cannot qualitatively reproduce the behavior of the {\em lines}. In
particular, they cannot produce the observed emission all of the time.
Thus, something important is missing from the calculations- perhaps
concerning the fate of shock waves propagating upwards into a magnetic
``canopy'', perhaps concerning different propagation modes (MHD
effects), or energetically non-connected material lying along the line of site (like magnetic flux-tubes). In any case this comparison verifies that SUMER can
indeed provide new information on the gas dynamics through observing
new spectral features.

\acknowledgements 

Philip Judge is very grateful to NORDITA, and the
Institute of Theoretical Astrophysics at the University of Oslo, for
support of a 6 month collaborative leave in Oslo in 1996, and to UCAR
for supporting that leave. We are very grateful to Philippe Lemaire
for providing the \ion{O}{4} spectroheliogram.
The SUMER project is financially supported by DARA, CNES, NASA and
 PRODEX (Swiss contribution).

\newpage

\begin{table*}
\caption{SELECTED SUMER OBSERVATIONS FROM 12 MARCH 1996\label{tab:observations}}
\protect\begin{tabular}{lllll}

Item & Start UT & End UT & Exp. & Target lines \\

\ldots & 09:14:04 & 10:11:32 & 15 & \ion{O}{4} 790.112+790.199
(spectroheliogram) \\
1 & 10:31:26&11:31:26 & 15 & \ion{O}{1} 1358.51, (\ion{O}{1}+\ion{C}{1}) 1355.72, \ion{C}{1} 1364.16\\
2 & 11:31:55&12:31:55 & 15 & \ion{N}{1} 1319.00, \ion{C}{1} 1329.58, \ion{C}{2} 1334.53 \\
3 & 12:33:06& 13:33:07& 15 & \ion{N}{1} 1199.55, Si II 1196.39, Si III 1206.51 \\
4 & 13:33:34&14:33:55 & 20 & \ion{C}{3} 1175.71, He I 584.334, \ion{C}{1} 1156.03, \ion{O}{1} 1152.15\\

\end{tabular}
\end{table*}

\begin{figure*}[p] 
 \caption[]{
 SUMER data showing the intensity profile as a function of
 wavelength in Doppler units ($x$-axis, units of km~s$^{-1}$) at
 spatial positions $x=95-96$ along the $1\times 120$ arcsecond
 slit, and time ($y$-axis, units of seconds). Data for two
 separate timeseries are shown, the 1355\AA\ region in the upper
 panel, the 1319\AA\ region in the lower panel. The ``grains'' are
 seen as brightenings with an intermittent $\sim$3 minute
 periodicity and a characteristic spectral profile: brightenings
 are accompanied by a net blue shift of $\sim 5$km~s$^{-1}$. The
 $x$-axes have not been corrected for zero point offsets or long
 term drifts. Telemetry gaps occurred
 near 2800 and 3300 s (lower panel).
 \label{fig:fig_qs} 
 }
\end{figure*}

\begin{figure*}[p] 
 \caption[]{
 Continuum intensity ({\it left}), total line intensity
 (continuum intensity subtracted)({\it middle}), line
 Doppler shift ({\it right})
 as function of position along the slit ($x$-axis) and
 time ($y$-axis) for the \ion{N}{1} 1319 line ({\it top})
 and the \ion{C}{2} 1334 line ({\it bottom}). The continuum
 intensity is given in counts ({\it top left}) and as the
 corresponding radiation temperature ({\it bottom left}).
 Doppler shifts are shown with upward velocity (blue-shift) bright.
 All data from the same timeseries.
 \label{fig:fig_moments_qs} 
 }
\end{figure*}

\begin{figure}[htbp]
 \caption[]{
 The behavior with time of intensity and Doppler shift of the
 continuum, \ion{N}{1} and \ion{C}{2} lines. The plots are summed
 over spatial pixels 93-97.
 \label{fig:fig_timeser_qs} 
 }
\end{figure}

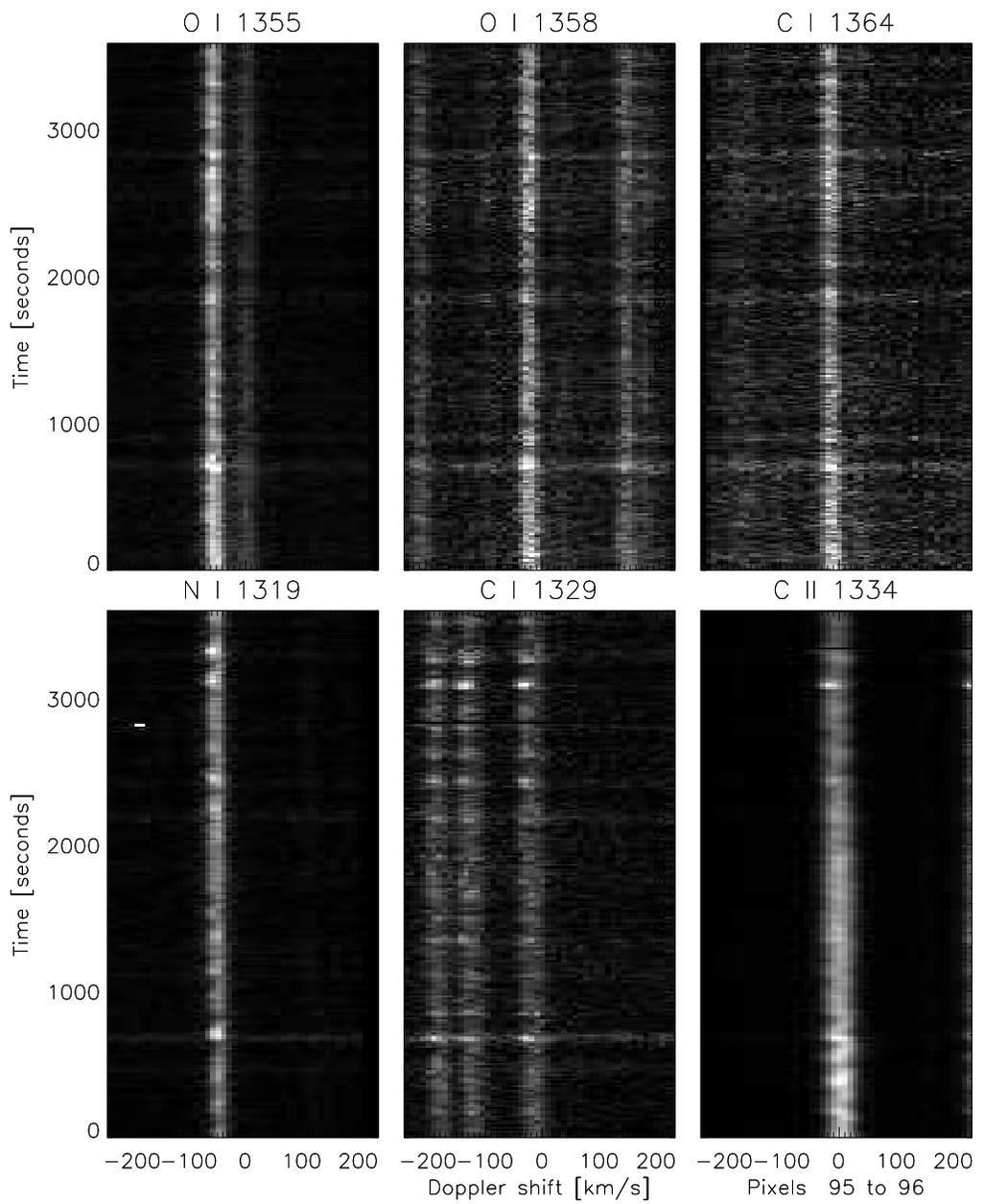
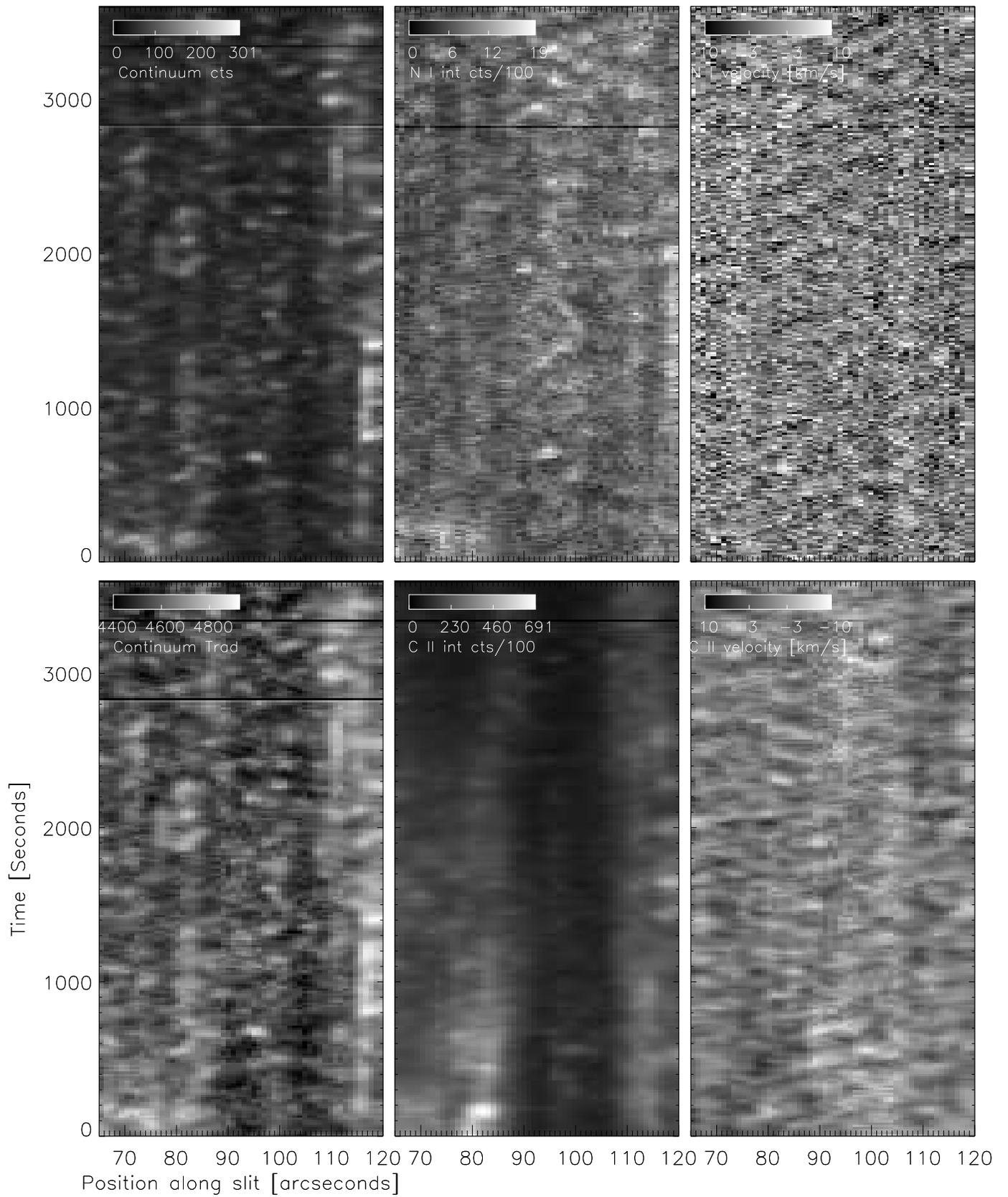
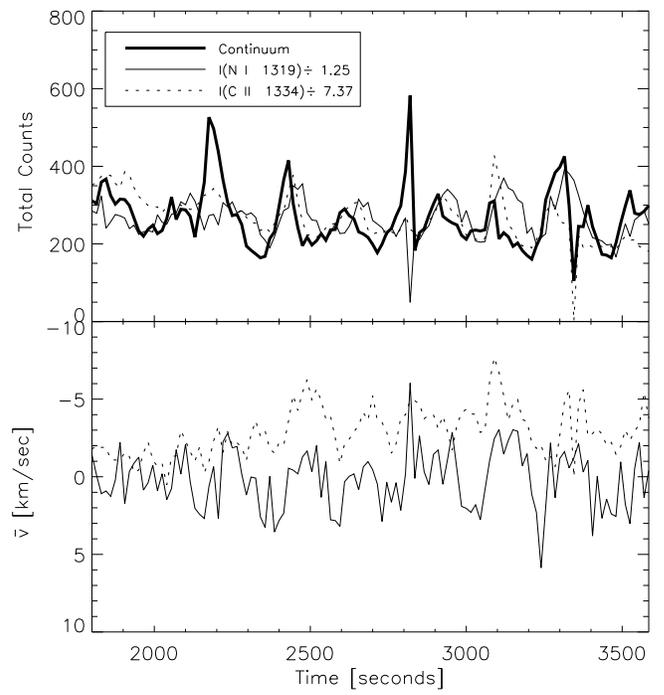
\end{document}